\documentclass[reprint, amsmath,amssymb, aps]{revtex4-2}

\usepackage{graphicx}
\usepackage{dcolumn}
\usepackage{bm}
\usepackage{physics}
\usepackage{xcolor}
\usepackage{comment}
\usepackage{fancyhdr}
\def\black#1{{\color{black} {#1}}}

\begin{document}
\fancyhead[R]{\ifnum\value{page}<2\relax\else\thepage\fi}
\title{Quantum Walk Inspired Dynamic Adiabatic Local Search}
\author{Chen-Fu Chiang}
 \affiliation{Department of Computer Science, State University of New York Polytechnic Institute, Utica, NY 13203 USA}
 \email{chiangc@sunypoly.edu}
\author{Paul M. Alsing}
\affiliation{ Information Directorate, Air Force Research Laboratory, Rome, NY 13441, USA}
\email{corresponding author: paul.alsing@us.af.mil}

\date{\today}

\begin{abstract}
We investigate the irreconcilability issue that raises in translating the search algorithm from the Continuous-Time Quantum Walk (CTQW) framework to the Adiabatic Quantum Computing (AQC) framework. 
\black{For the AQC formulation to evolve along the same path as the CTQW requires a constant energy gap in the former Hamiltonian throughout the AQC schedule. To resolve the issue, we modify the CTQW-inspired AQC catalyst Hamiltonian with a $Z$ oracle operator. Through simulation we demonstrate that the total running time for the proposed approach remains optimal.} Inspired by this solution, we further investigate \black{adaptive scheduling for the catalyst Hamiltonian and its coefficient function in the adiabatic path to improve the adiabatic local search.}
\end{abstract}
\maketitle
\thispagestyle{fancy}  

\section{Introduction}
Quantum technologies have advanced dramatically in the past decade, both in
theory and experiment. \black{From the view of theoretical computational complexity, 
Shor’s factoring algorithm \cite{shor1994algorithms} and Grover’s search algorithm  \cite{grover1996fast} are 
well-known for their improvements over the best possible classical algorithms designed for the same purpose.}
%
%
From a perspective of universal computational models, Quantum Walks (QWs) have become a
prominent model of quantum computation due to their direct relationship to the
physics of the quantum system \cite{farhi1998quantum, kempe2003quantum}. It has been
shown that the QW computational framework is universal for quantum computation
\cite{childs2009universal, lovett2010universal}, and many algorithms now are
presented directly in the quantum walk formulation rather than through a circuit
model or other abstracted method \cite{farhi1998quantum, qiang2016efficient}.
Besides being search algorithms,
CTQWs have been applied in fields such as quantum
transport\cite{caruso2009highly, mohseni2008environment,
rebentrost2009environment,plenio2008dephasing}, state transfer
\cite{bose2003quantum, kay2010perfect}, link prediction in complex networks
\cite{omar2019quantum} and the creation of Bell pairs in a random network
\cite{chakraborty2016spatial}.
Some other well-known universal models include the quantum circuit model \cite{shor1998quantum, yao1993quantum,
jordan2012quantum}, topological quantum computation \cite{nayak2008non},
adiabatic quantum computation (AQC) \cite{mizel2007simple}, resonant transition based
quantum computation \cite{chiang2017resonant} and measurement based quantum
computation \cite{morimae2012blind, gross2007novel, briegel2009measurement,
raussendorf2003measurement}. 
\black{Each model might has its own bottleneck. 
Investigation on the relationship among the frameworks 
helps identify the violations when mapping frameworks and potential solutions.
By studying the mapping, one can extend the techniques from one framework to another
for some potential speedup \cite{Cutugno:2022}}.

\black{
In this work we investigate the irreconcilability issue that arises when translating the search algorithm from
the Continuous-Time Quantum Walk (CTQW) framework to the Adiabatic Quantum Computing
(AQC) framework as first pointed out by Wong and Meyer \cite{wong2016irreconcilable}. This irreconcilability issue can be described as follow. One first notes that the CTQW is the unique continuous-time quantum walk formulation of Grover’s discrete search algorithm. While the CTQW search evolves the initial unbiased (equal amplitude) state to the unknown (marked) state on the order of time $T\sim \mathcal{O}(\sqrt{N})$
(where $N$ is the size of search space), it does not follow the same evolution path (on the Bloch sphere) as that of Grover's  algorithm. The uniqueness of the CTQW formulation stems from the fact that the unknown marked state only acquires a (time-dependent) phase from the oracle operation. Most importantly the marked states does not undergo evolution, and thus the CTQW effective employes a dichotomous ``Yes/No" oracle, for which the discrete Grover's algorithm has been proven to be optimal. 
}

\black{
The AQC formulation of the search algorithm with a non-uniform adiabatic evolution schedule \cite{roland2002quantum} also finds the marked state in time $T\sim \mathcal{O}(\sqrt{N})$ while at the same time following the same path as Grover's algorithm. Thus, if one investigates  what adiabatic Hamiltonian gives rise to the same evolution path as the CTQW formulation, one finds \cite{wong2016irreconcilable} the AQC formulations introduces an extra ``catalyst" Hamiltonian which introduces structure beyond the standard ``Yes/No" oracle employed in the CTQW or discrete (Grovers) search algorithm. A scaled version of the AQC Hamiltonian leads to a constant energy gap that implies that the  marked state can be found in time 
$T\sim \mathcal{O}(1)$.
This discrepancy between the formulations of the two versions of a continuous time search algorithm  was termed the ``irreconcilability (difference) issue" between CTQW and AQC by Wong and Meyer \cite{wong2016irreconcilable}.
}


\black{In this work we address the CTQW/AQC search algorithm irreconcilability issue by modifying the constant energy gap Hamiltonian of the AQC formulation. Our contribution is twofold}. 
We first adapt the result from the mapping of CTQW to AQC by 
selecting the regular oracle $Z$ operator as the catalyst Hamiltonian and explore an alternative for the 
coefficient function for the catalyst Hamiltonian in order to attempt to avoid the irreconcilability issue. Through the simulation, 
the modified model provides optimal results in terms of time required for search. 
\black{We then apply this modification to adiabatic local search by adding an additional sluggish parameter $\delta$ which delineates the width of the adiabatic run time schedule over which the catalyst Hamiltonian effectively acts (i.e. the ``slowdown" region in the vicinity of the system's smallest energy gap $\Delta$).
The sluggish parameter tracks the increase of running time $t=t(s)$ with respect to 
schedule parameter $0\le s\le 1$  where $\delta=|d^2t/ds^2|$. The catalyst is employed when $\delta \ge\delta_0$ to facilitate the process, where we have found that the threshold value of $\delta_0=64$ provides good results.} 

The outline of this work is as follows. The background information regarding CTQW and AQC is given in section \ref {sect:background} where the translation of CTQW to AQC is described in section \ref{sect:CTQW-AQC}. The irreconcilability issue that occurs during the translation is explained in section \ref{sect:irrecon} and our proposed solution is provided in section \ref{sect:m-qw-aqc}. The mapping of Grover search to AQC as an adiabatic local search is summarized in section \ref{sect:adpt-ags}. We propose and describe the catalyst Hamiltonian mechanism in section \ref{sect:catrelease} and determine the sluggish interval  where it is employed. We further explore three coefficient functions of the catalyst Hamiltonian in section \ref{sect:catcoefficient}. The simulation results for proposed modifications are discussed in section \ref{sect:experiment}. Finally, our conclusions are  given in section \ref{sect:conclude}.

\section{Background}\label{sect:background}

\subsection{Continuous-Time Quantum Walk}\label{sect:ctqw}
Given a graph $G=(V, E)$, where $V$ is the set of vertices and $E$ 
is the set of edges, the CTQW on $G$ is defined as follows. 
Let $A$ be the adjacency matrix of $G$, the $|V| \times |V|$ 
matrix is defined component-wise as
\begin{equation}
A_{ij} =
  \begin{cases}
   1 & \text{if } (i,j) \in E, \\
   0 & \text{otherwise}
  \end{cases}
\end{equation}
where $i, j \in V$. A CTQW starts with a uniform superposition state $\ket{\psi_0}$ in the
space, spanned by nodes in $V$, evolves according to the Schr\"odinger equation
with Hamiltonian $A$.
After time $t$, the output state is thus
\begin{equation}
\ket{\psi_{t}} = e^{-iAt}\ket{\psi_0}.
\end{equation}
\newline The probability that the walker is in the state
$\ket{\tau}$ at time $t$ is given by $|\bra{\tau}{e^{-iAt}\ket{\psi_0}}|^2$.
To find the marked node $\ket{\omega}$ starting from an
initial state $\ket{\psi_0}$ via a CTQW, one has to maximize the success probability
\begin{equation} |\mel{\omega}{e^{-iAt}}{\psi_0}|^2  \end{equation} while minimizing the
time $t$. For instance, initially at time $t = 0$, the success probability is
\begin{equation}
|\mel{\omega}{e^{-iA0}}{\psi_0}|^2 = O(\frac{1}{|V|}).
\end{equation}
The success probability is extremely small when the 
search space $|V| = N $ is large and $\ket{\psi_0}$ is
a uniform superposition state.  

When applied to spatial search, the purpose of a CTQW
is to find a marker basis state $\ket{\omega}$\cite{childs2004spatial,childs2003exponential}.
For this purpose, the CTQW starts with the initial state
$\ket{\psi_0} =\sum_{i=1}^{N}\frac{1}{\sqrt{N}}\ket{i}$,
and evolves according to the Hamiltonian\cite{novo2015systematic}
\begin{equation}\label{eqn:CTQW_gamma}
 H = -\gamma A - \ket{\omega}\bra{\omega}
\end{equation}
where $\gamma$ is the coupling factor
between connected nodes. \black{The value of $\gamma$
has to be determined based on the graph structure} 
such that the quadratic speedup of CTQW can be preserved. Interested readers can refer to
\cite{childs2004spatial, novo2015systematic} for more details.

\subsection{Adiabatic Quantum Computing} \label{sect:aqc}
In the AQC model, $H_0$ is the initial Hamiltonian, $H_f$ is the final
Hamiltonian. The evolution path for the time-dependent Hamiltonian is
\begin{equation}\label{eqn:aqc}
H(s) = (1-s)H_0 + sH_f
\end{equation}
where $0 \leq s  \leq 1$ is a schedule function of time $t$. For convenience, we 
denote $s$ as $s(t)$ and use them interchangeably. 
The variable $s$ increases slowly enough such that the initial ground state evolves and remains as the instantaneous ground state of the system. More specifically,
\begin{align}
H(s(t))\ket{\lambda_{k,t}} = \lambda_{k,t}\ket{\lambda_{k,t}}
\end{align}
where $\lambda_{k,t}$ is the corresponding eigenvalue the eigenstate
$\ket{\lambda_{k,t}}$ at time $t$ and $k$ labels for the $k_{th}$ excited eigen-state. 
The minimal eigenvalue gap is defined as
\begin{align}
g = \min_{0 \leq t \leq Ta}(\lambda_{1, t} - \lambda_{0,t})
\end{align}
where $T_a$ is the total evolution time of the AQC. Let $\ket{\psi (T_a)}$ be
the state of the system at time $T_a$ evolving under the Hamiltonian $H(s(t))$
from the ground state $\ket{\lambda_{0,0}}$ at time $t=0$. The Adiabatic theorem
\cite{farhi2000quantum, albash2018Adiabatic} states that the final state
$\ket{\psi (T_a)}$ is $\epsilon$-close to the real ground state
$\ket{\lambda_{0,T_a}}$ as
\begin{align}\label{eqn:aqc_limit_approx}
|\braket{\lambda_{0, T_a}}{\psi(T_a)}|^2 \leq 1 - \epsilon^2,
\end{align}
provided that
\begin{align}
\frac{|\bra{\lambda_{1,t}}\frac{dH}{dt}\ket{\lambda_{0,t}}|}{g^2} \leq \epsilon.
\end{align}

There are several variations of AQC to improve the performance. The variations
are based on modifying the initial Hamiltonian and the final Hamiltonian
\cite{10.5555/2011395.2011396, perdomo2011study} or adding a catalyst
Hamiltonian $H_e$ \cite{10.5555/2011395.2011396}, which is turned on/off at the beginning/end of the adiabatic evolution. 
In this work, we are
interested in the catalyst approach.  
A conventional catalyst Hamiltonian assisted AQC path is expressed as
\begin{align}\label{eqn:typical_catalyst_aqc}
H(s) = (1-s)H_0 + s(1-s)H_e + sH_f.
\end{align}

\section{Continuous Time Quantum Walk to Adiabatic Search Mapping} \label{sect:CTQW-AQC}
\noindent 
One can construct a time-dependent AQC Hamiltonian $H(s)$ as shown in \cite{wong2016irreconcilable} where the Adiabatic search follows the CTQW search on a complete graph with $N$ vertices. Let us define the following variables. The coupling factor $\gamma$ is set to $1/N$ and $\ket{\psi_0}$ is the uniform superposition of all states in the search space. State $\ket{r}$ is the uniform
superposition of non-solution states, state $\ket{\omega}$ is the solution state. 
Treating the state evolving in CTQW system as the time-dependent ground state of $H(s)$,
one constructs $H(s)$ in the $\{\ket{\omega},  \ket{r}\}$ basis as \cite{wong2016irreconcilable}
\begin{align}\label{eqn:ctqw_aqc}
H(s) =& \sqrt[4]{\frac{s(1-s)}{4\epsilon^2N}}[(1-s)H_0 + \sqrt{s(1-s)}H_e + s H_f]
\end{align}
where $s(t) = sin^2(\frac{t}{\sqrt{N}})$ with
\black{
\begin{align}\label{eqn:h0hf}
H_0 &= \dyad{\psi_0^{\perp}}{\psi_0^{\perp}} - \dyad{\psi_0}{\psi_0}, \quad  H_f =\dyad{\gamma} - \dyad{\omega}{\omega}, \nonumber \\
H_e &= 2i\sqrt{\frac{N-1}{N}}( \dyad{r}{\omega} - \dyad{\omega}{r}),
\end{align}
or explicitily in the $\{\ket{w}, \ket{r}\}$ basis as
\begin{align} \label{eqn:HamiltoniansCTQW}
H_0  &=
\begin{pmatrix}
\frac{N-2}{N} & -2\frac{\sqrt{N-1}}{N} \\
-2\frac{\sqrt{N-1}}{N} & -\frac{N-2}{N} \\
\end{pmatrix},  \\
H_e & =
\begin{pmatrix}
0 & -2i\sqrt{\frac{N-1}{N}} \\
2i\sqrt{\frac{N-1}{N}}  & 0\\
\end{pmatrix},\;
H_f  =
\begin{pmatrix}
-1 & 0 \\
0  & 1\\
\end{pmatrix}. \nonumber
\end{align}
}
\subsection{The Irreconcilability Issue: Constant Gap Catalyst Hamiltonian and Small Norm} \label{sect:irrecon}
\noindent 
The main concerns that are raised from Eqn. (\ref{eqn:ctqw_aqc}) are twofold. 
The first issue is the factor $ \sqrt[4]{\frac{s(1-s)}{4\epsilon^2N}}$ of $H(s)$. 
The adiabatic theorem \cite{griffiths2018introduction} states 
the system achieve a fidelity of $1-\epsilon$ to the target state, provided that 
\begin{equation}\label{eqn:dhdtgmin}
\frac{|\langle\frac{dH}{dt}\rangle_{0,1}|}{g_{min}^2} \leq \epsilon \text{,  where } g_{min} = \min_{0\leq t \leq T}E_1(t) - E_0(t) .
\end{equation}
Here $\langle\frac{dH}{dt}\rangle_{0,1}$ are the matrix elements of $dH/dt$ between the two corresponding eigen-states. 
$E_0(t)$ and $E_1(t)$ are the ground energy and the first excited energy of the system at time $t$. Given the $H(s)$ in 
 Eqn. (\ref{eqn:ctqw_aqc}), one might conclude that a factor of  $ O(\sqrt[4]{1/N})$ significantly reduces the required time to achieve $1-\epsilon$ precision.  This might be misleading as the $g_{min}$ of $H(s)$ also carries the same factor. 
The second issue is that the catalyst $H_e$ provides power
greater than a typical Yes$\slash$No oracle as it maps non-solution states 
to a solution state and a solution state to non-solution states. 
Provided initially the we start with a superposition state with amplitude of
 $\sqrt{\frac{N-1}{N}}$ for a non-solution, it takes time of $O(1)$ for this catalyst  
to drive the initial (unbiased, equal amplitude) state
to the solution state. 
In the following we will relax this constraint by using a normal oracle. 
For the rest of the paper, let us simply treat $\epsilon\ll 1$ as some small negligible constant. 

\subsection{Modified CTQW-Inspired Adiabatic Search} \label{sect:m-qw-aqc}
\noindent 
In Eqn.(\ref{eqn:ctqw_aqc}), the following parameters were computed during the mapping \cite{wong2016irreconcilable}: 
\begin{itemize}
    \item the scaling factor $ \sqrt[4]{\frac{s(1-s)}{4\epsilon^2N}}$\; of  Hamiltonian\, $H_0$, 
    \item $H_e = 2i\sqrt{\frac{N-1}{N}}( \dyad{r}{\omega} - \dyad{\omega}{r})$, catalyst Hamiltonian 
    \item the coefficient function of $H_e$ as $\sqrt{s(1-s)}$.  
\end{itemize}
In \cite{aharonov2008Adiabatic} the cost of the adiabatic algorithm was defined to be the dimensionless quantity 
(using $\hbar=1$)
\begin{equation}\label{eqn:Adiabatic_cost}
cost = t_f \max_{s}||H(s)||, 
\end{equation}
where $t_f$ is the running time. To prevent the cost from being manipulated to be arbitrarily small by changing the time units, or distorting the scaling of the algorithm by multiplying the Hamiltonians by some size-dependent factor as shown in the irreconcilability concern \cite{wong2016irreconcilable}, the norm of $H(s)$ should be fixed to some constant, such as 1. \\

\noindent
To address the irreconcilability issue, the scaling factor is dropped and the catalyst Hamiltonian $H_e$ is modified. Since $H_e = 2\sqrt{\frac{N-1}{N}} i XZ$ in the $\{\ket{\omega},  \ket{r}\}$ basis 
provides more power than a standard Oracle, for our modification we remove the imaginary number $i$ and the $X$ operator. The operator $Z$ alone behaves as a conventional ``Yes \slash No" oracle in the $\{\ket{\omega},  \ket{r}\}$ basis. 
Let $M=2\sqrt{\frac{N-1}{N}}$ and choose the modified adiabatic path $H_m(s)$ as
\begin{align}\label{eqn:ctqw_m_aqc}
H_m(s) =&(1-s)H_0 + f_{z}(s)MZ + s H_f,
\end{align}
where $f_{z}(s)$ is our chosen $s$-dependent coefficient for catalyst $Z$. 
In addition to $f_{z}(s) = \sqrt{s(1-s)}$ that was used in \cite{wong2016irreconcilable}, 
functions that reach its maximum when $s=1/2$ are good candidates for $f_{z}(s)$, such as $f_{z}(s)=\frac{\sin(s \pi)}{2}$. 
\black{The use of the factor $1/2$ on the sine function is to offset the magnitude $M$ 
to bound the norm of $H_e$ as described in Eqn. (\ref{eqn:Adiabatic_cost}).}

\section{Grover Search to Adiabatic Local Search Mapping}\label{sect:adpt-ags}
\begin{figure}[htbp]
\centerline{\includegraphics[scale=.25]{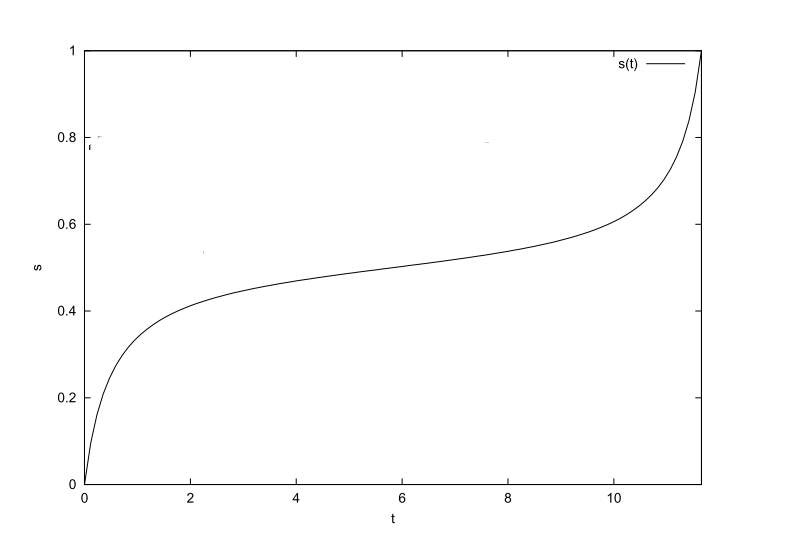}}
\caption{ Schedule $s$  in terms of time $t$ with $N=64$ in adiabatic local search, as observed in \cite{roland2002quantum}.}
\label{fig:Adiabaticgrover_org}
\end{figure}

In this section we consider the mapping of Grover's algorithm to an adiabatic search.
Given the initial driving Hamiltonian $H_0$ and the final Hamiltonian $H_f$ as 
\black{
\begin{equation}
 H_0 =I - \dyad{\psi_0}{\psi_0}, \quad H_f =I - \dyad{\omega}{\omega}, 
\end{equation}
where 
\begin{align} \label{eqn:HamiltoniansAGS}
H_0  =
\begin{pmatrix}
\frac{N-1}{N} & -\frac{\sqrt{N-1}}{N} \\
-\frac{\sqrt{N-1}}{N} & \frac{1}{N} \\
\end{pmatrix},  
H_f  =
\begin{pmatrix}
0 & 0 \\
0  & 1\\
\end{pmatrix},
\end{align}
}
in the $\{\ket{\omega},  \ket{r}\}$ basis. The adiabatic path \cite{roland2002quantum, wong2016irreconcilable} in the $\{\ket{\omega}, \ket{r}\}$ basis is given by
\begin{align}\label{eqn:grover_hs}
H(s) &= (1-s)H_0 + sH_f \\
&=
\begin{pmatrix}
(1-s)\frac{N-1}{N} & -(1-s)\frac{\sqrt{N-1}}{N} \\
-(1-s)\frac{\sqrt{N-1}}{N} & 1 - (1-s)\frac{N-1}{N} \\
\end{pmatrix}.
\end{align} 
Instead of employing a linear evolution of $s(t)$, 
Eqn.(\ref{eqn:grover_hs})  adapts the evolution $ds/dt$ to the local adiabaticity condition \cite{roland2002quantum} such that 
\begin{equation}\label{eqn:dsdt}
|\frac{ds}{dt}| = \epsilon g^2(t) 
\end{equation} 
where $g(t)$ is the energy gap of the system at time $t$. The running time $t$ is then a function of schedule $s$ such that 
\begin{align}
t(s) &= \frac{N}{2\epsilon \sqrt{N-1}}\Big\{\arctan(\sqrt{N-1}(2s-1)) \\ 
&+ \arctan(\sqrt{N-1})\Big\}.
\end{align}
The relation between the schedule $s$ and the running time 
$t$ is shown in Figure \ref{fig:Adiabaticgrover_org}. It is clear that the system evolves quickly when the gap is large ($s$ away from $1/2$) and slowly when the gap is small ($s \simeq 1/2$) \cite{roland2002quantum}. In this example, the sluggish period $s \in [0.4, 0.6]$. For completeness, we provide the formal proof of the close form of the squared gap function  $g^2(t)$ 
(second order in $s$) with respect to the schedule $s$ in Appendix \ref{sect:sTot}.

\subsection{Adaptive Scheduling }\label{sect:qw-ags}
\noindent 
\black{For a fixed schedule of a adiabatic path, the schedule $s$ moves fast when the eigen-energy gap is large, and slowly when the gap is small. We desire to employ the catalyst Hamiltonians $H_e$ to amplify the eigen-energy gap during the ``slow down" period such that the total time to pass through the sluggish period is reduced ($s \in [0.4, 0.6]$ in Fig.(\ref{fig:Adiabaticgrover_org})).} 

\subsubsection{Schedule Dependent Gap Function}\label{sect:gapmin}
In this section, we consider employing  gap-dependent scheduling functions.
Let $H_f$ be an arbitrary 2 by 2 Hermitian Hamiltonian. Let the time-dependent  Hamiltonian $H(s)$ be
\begin{equation}\label{time_dep_H}
H(s) = (1-s) H_o + f_x(s) \sigma_x + f_z(s)\sigma_z + sH_f.
\end{equation}
Operators $\sigma_x$ and $\sigma_z$ are chosen as catalyst Hamiltonians. Let  
$H_o = 
\begin{bmatrix}
a & c \\
c & b \\
\end{bmatrix},
H_f = 
\begin{bmatrix}
p & r \\
r & q \\
\end{bmatrix}
$ where $a, b, c, p, q, r$ are some given constants. The matrix form of the time-dependent 
Hamiltonian is given by
\begin{equation}
H(s) = 
\begin{bmatrix}
(1-s)a +sp +f_z(s)& (1-s)c+sr+f_x(s) \\
(1-s)c+sr+f_x(s) & (1-s)b +sq-f_z(s) \\
\end{bmatrix}
\end{equation}
and the schedule-dependent gap can be analytically computed to yield 
\begin{align}\label{eqn:gminsqr}
g^2(s) &= ((1-s)(a-b)+s(p-q) +2f_z(s))^2  \nonumber \\
       &+ 4( (1-s)c+sr+f_x(s) )^2,
\end{align}
(see  see Appendix \ref{sect:simpleAmp} for a derivation).
By using Eqn. (\ref{eqn:dsdt}),  the total running time $T_{strt}^{stp}$ from $s = s_{strt}$ to $s= s_{stp}$  
 is thus
\begin{equation}\label{eqn:totalTimestrtstp}
T_{s_{strt}}^{s_{stp}}  =  \int_{s_{strt}}^{s_{stp}} \frac{ds} {\epsilon g^2(s)}
\end{equation}
where $0 \leq s_{strt} \leq s_{stp} \leq 1$. \black{In brief, the time spent during a certain period of a schedule can be obtained by use of gap function. The gap function can be expressed via the entries of $H_0$, $H_e$, $H_f$, schedule $s$ and the coefficient functions of the catalyst Hamiltonians.}

\subsubsection{Determining the Sluggish Interval \\ for the Catalyst Hamiltonian}\label{sect:catrelease}
\begin{figure}[htbp]
    \centering
    \includegraphics[width=0.30\textwidth]{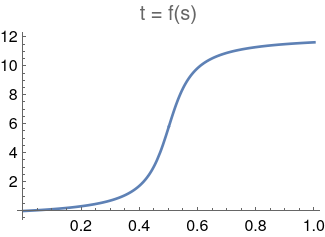} 
    \caption{Time $t$ as a function of schedule $s$ for adiabatic local search with $N=64$.}
    \label{fig:timeschedule_1}
\end{figure}

\begin{figure}[htbp]
    \centering
    \includegraphics[width=0.30\textwidth]{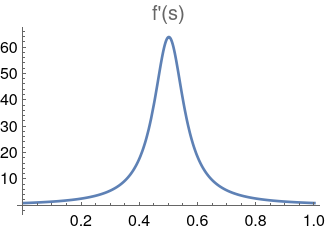}
    \caption{$dt/ds$ for adiabatic local search with $N=64$.}
    \label{fig:timeschedule_2}
\end{figure}

\noindent 
By using the condition $f'(s) = dt/ds = \frac{1}{\epsilon g^2(s)}$ (see Appendix \ref{sect:sTot}), the region
where the gap quickly significantly decreases or increases is during the sluggish period of $s$. 
That is the portion of schedule the $s$ where catalyst should be employed. The region where 
$|df^2(s)/ds^2)| \geq \delta_0$ is the sluggish period. \black{The threshold value  $\delta_0=64$ was chosen because if we choose a threshold proportional to $N$, as $N$ increases exponentially, the quantity $d^2t/ds^2$ might never reach the $N$-dependent threshold within  the adiabatic evolution schedule $0\le s\le 1$.} By using this threshold, the starting point $s_{strt}^{slug}$ and the stopping point  $s_{stp}^{slug}$ used to mark the sluggish period can be identified. Using the example in  \cite{roland2002quantum}, 
we can re-plot and get $t$ as a function of $s$ as $t = f(s)$ and $f'(s) = dt/ds$ in Figure \ref{fig:timeschedule_1} - \ref{fig:timeschedule_2} with $N=64$. 

\subsubsection{Catalyst Coefficient Functions}\label{sect:catcoefficient}
\noindent 
\black{As discussed in section  \ref{sect:m-qw-aqc}, we are interested in the $H_e = Z$ case in Eqn.(\ref{eqn:ctqw_m_aqc}) and its coefficient function $f_z(s)$. Three coefficient functions of the catalyst Hamiltonian $Z$ are proposed as the following}
\begin{align}\label{adapt_impv_func}
f_{z}^{sine} (s) &= \sine (((s-s_{strt}^{slug})* \pi)/(s_{stp}^{slug} - s_{strt}^{slug})), \\
f_{z}^{ss} (s) & = (s-s_{strt}^{slug})(s_{stp}^{slug} -s), \nonumber \\
f_{z}^{grid} (s) &= a *f_{z}^{sine} (s)  +  b*(f_{z}^{sine} (s) )^2 \nonumber 
\end{align}
where $ 0 \leq a, b \leq 1$ under the constraint that  $a^2+b^2=1$.  In the grid search $a$ increased from 0 to 1 by $0.1$ in each iteration. From the 10 pairs of
$(a,b)$, we find the values of $a, b$ that give the shortest sluggish time interval.  

\section{Experiment \& Result }\label{sect:experiment}
For our simulations we used (Wolfram) Mathematica (version 12.3 run on a Linux Ubuntu 20.04 LTS laptop). The code is available upon request. 
The running time is based on Eqn.(\ref{eqn:totalTimestrtstp}).  
The size $N$ (number of nodes) ranges from $2^5, 2^6, \cdots$ to $2^{25}$. We observe the corresponding 
running time and sluggish time for each of the proposed models. \black{The result of the original adiabatic local search serves as the baseline for comparison, which used $N=64$ \cite{roland2002quantum}.} 
In this work, we generalize the setting for any arbitrary size $N$.

\black{Given an arbitrary complete graph of size $N$ with coupling factor $1/N$, one can compute the entries in the reduced Hamiltonian for $H_0$ and $H_f$ in the $\{\ket{\omega},  \ket{r}\}$ basis.  
The values of variables $a, b, c, p, q$ and $r$ as discussed in section \ref{sect:gapmin} can be obtained from Eqn.(\ref{eqn:HamiltoniansCTQW}) for the CTQW case, and from Eqn.(\ref{eqn:HamiltoniansAGS}) for the adiabatic local search. It is worth noticing that the ground state energy is 
$-1$ in the CTQW case, but is $0$ in the adiabatic local search case. 
Based on the adiabatic path Eqn.(\ref{time_dep_H}), and the gap function in Eqn. (\ref{eqn:gminsqr}) with given schedule $s$, coefficient function $f_z(s)$ for $\sigma_z$, we perform the simulation with the running time computed from Eqn.(\ref{eqn:totalTimestrtstp}).} 

\subsection{Modified CTQW-Inspired Adiabatic Search Simulation} \label{sect:m-qw-aqc-sim}
\noindent 
This experiment aimed to demonstrate that the modified adiabatic paths addressing the irreconcilable issues 
remain optimal. The three proposed modifications we explored are as follows:  

\begin{itemize}
    \item \black{$H_{org}(s)$ takes Eqn. (\ref{eqn:ctqw_aqc}) and drops the scaling factor as explained in section \ref{sect:m-qw-aqc}.} The adiabatic path is 
    $H_{org}(s) = (1-s)H_0 + \sqrt{s(1-s)}H_e + s H_f$
    \item $H_{m1}(s)$ replaces the computed catalyst Hamiltonian $H_e$ with an ordinary $Z$ oracle operator and keeps the magnitude $M$. This was used to address the constant gap $H_e$ irreconcilability issue. We have
    \newline $H_{m1}(s) = (1-s)H_0 + \sqrt{s(1-s)}MZ + s H_f$
    \item $H_{m2}(s)$ uses $\frac{\sin(s \pi)}{2}$ as the coefficient function for the catalyst Hamiltonian $Z$. The adiabatic path is 
    $H_{m2}(s) = (1-s)H_0 + \frac{\sin(s \pi)}{2}MZ + s H_f$
\end{itemize} 
\black{For the above three models, simulations were run on Hamiltonian of size $N \in [2^5, 2^{6}, \cdots, 2^{25}]$. In the following figures, the abscissa is $\log_2 N$ while ordinate is the required total running time $T$. The time is computed based on 
Eqn.(\ref{eqn:totalTimestrtstp}). As the dimension of the Hamiltonian increases, the difference in running times for the three models considered are magnified.}  
\begin{figure}[htbp]
    \centering
    \includegraphics[width=0.55\textwidth]{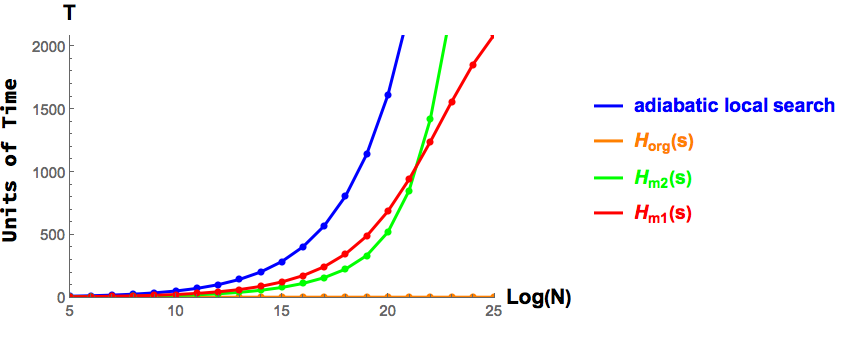}
    \caption{Case when $N \in [2^5, 2^{25}]$ and the running times of $H_{org}(s)$ (orange), $H_{m1}(s)$ (red), and $H_{m2}(s)$ (green) with the original adiabatic local search (blue) serving as the baseline. }
    \label{fig:ctqw-aqc-m-comparison-N25}
\end{figure}

\noindent
The simulation results are shown in Figure \ref{fig:ctqw-aqc-m-comparison-N25}. 
\black{It is clear to see that $H_{org}$ is a constant time scheme as it does not scale as the size $N$ increases. 
This indicates the original catalyst Hamiltonian $H_e = MXZ$ in $H_{org}(s)$ indeed is a constant gap Hamiltonian. This also shows the irreconcilability issue as suggested in \cite{wong2016irreconcilable}.} 
From the simulations we can conclude that both $H_{m1}(s), H_{m2}(s)$ perform optimally with respect to running time, 
\black{namely $T\sim\mathcal{O}(\sqrt{N})$}, similar to that of the original adiabatic local search but with a minor constant factor which can be ignored in the Big O notation. \black{As the simulation suggests, both modified CTQW-inspired approaches outperform the original adiabatic local search. When the $N \leq 2^{21}$, the $H_{m2}(s)$ outperforms $H_{m1}(s)$. When problem size $N$ is larger then $2^{21}$, $H_{m1}(s)$ is a better choice over $H_{m2}(s)$.}  

\subsection{Adaptive Adiabatic Local Search Simulation With Various Coefficient Functions} \label{sect:m-gs-aqc-sim}
\black{In the previous section \ref{sect:m-qw-aqc-sim}, the proposed modifications are optimal, \black{in the sense that $T\sim\mathcal{O}(\sqrt{N})$ up to a minor constant factor}. For further improvement, the adaptive scheduling scheme is applied.
The adiabatic path to be explored is therefore 
\begin{equation}
 H_{adapt}(s) = (1-s)H_0 + f(s)Z + s H_f
 \end{equation} 
 where $f(s) \in [f_{z}^{sine}, f_{z}^{ss}, f_{z}^{grid}]$ as seen in Eqn. (\ref{adapt_impv_func}). The catalyst Hamiltonian $Z$ operator is only employed during the sluggish period and hence $f(s) = 0 $ when $s \notin [s_{strt}^{slug}, s_{stp}^{slug}]$. The $H_0$ and $H_f$ are based on Eqn. (\ref{eqn:HamiltoniansAGS}).} As the catalyst is only employed within the sluggish period, to compare the performance of each proposed modification, one only needs to compute the running time within this period. 

\begin{figure}[htbp]
    \centering
    \includegraphics[width=0.5\textwidth]{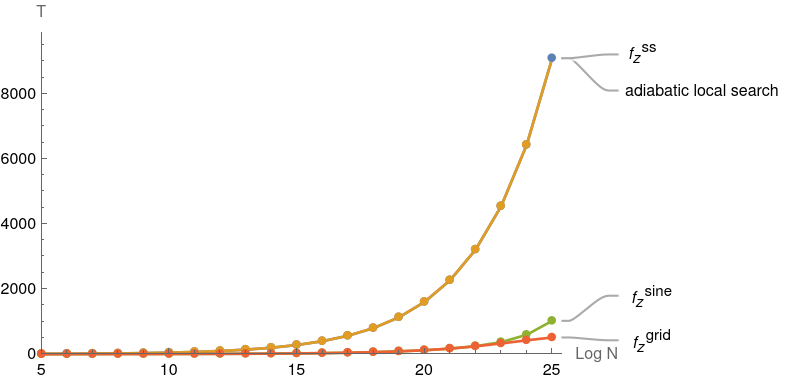}
    \caption{Case when $N \in [2^5, 2^{25}]$ and time spent in during the sluggish period for adiabatic paths with  
    ($f_{z}^{ss}, f_{z}^{sine}, f_{z}^{grid}$) coefficient functions where the original adiabatic 
    local search serves as the baseline.} 
    \label{fig:ags-m-comparison-small_1}
\end{figure}

\noindent
In Figure \ref{fig:ags-m-comparison-small_1},  $f_{z}^{ss}$  provides the minimal reduced sluggish time
while $f_{z}^{sine}$ and $f_{z}^{grid}$ provide significant improvements. The difference in the runtimes  becomes significant for $N\ge 2^{15}$. 
\begin{figure}[htbp]
    \centering
    \includegraphics[width=0.5\textwidth]{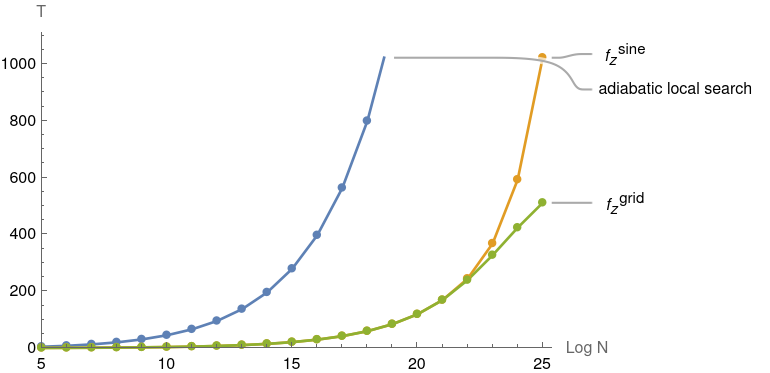}
    \caption{Case when $N \in [2^5, 2^{25}]$ and time spent in during the sluggish period for adiabatic paths with  
    ($f_{z}^{sine}, f_{z}^{grid}$) coefficient functions where the original adiabatic 
    local search serves as the baseline.} 
    \label{fig:ags-m-comparison-small_2}
\end{figure} 

 \noindent 
 In Figure \ref{fig:ags-m-comparison-small_2}, both $f_{z}^{sine}$ and $f_{z}^{grid}$ 
have a reduced sluggish time over $75\%$ in comparison to the original adiabatic local search when $N$ reaches $2^{25}$. 
$f_{z}^{sine}$ gradually outperforms the original adiabatic local search after $N=2^{10}$ and remains
almost as good as $f_{z}^{grid}$ till $N=2^{23}$. When $N =2^{25}$, the sluggish time of $f_{z}^{sine}$ is only twice that of $f_{z}^{grid}$. In general, the grid search is a costly procedure
as we have to run 10 pairs of $(a, b)$ for slightly different $H(s)$ for each value of $N=2^n$. If the time reduction of sluggish period is not greater than $90\%$ of the original, it might
be a better choice to use $f_{z}^{sine}$. For the near term it might be more beneficial to use $f_{z}^{sine}$ model, instead of the grid search model $f_{z}^{grid}$.

\section{Conclusion}\label{sect:conclude}
In this work, we investigated different Hamiltonians for
resolving the irreconcilability issue \cite{wong2016irreconcilable} when mapping the CTQW search algorithm to AQC. We modified the 
time-dependent Hamiltonian by (1) removing the original scaling CTQW factor $\sqrt[4]{\frac{s(1-s)}{4\epsilon^2N}}$, and (2) replacing $i\,X\,Z\to Z$  in the original catalyst $H_e$ Hamiltonian obtained from mapping CTQW to AQC. These modification were made in order to resolve the irreconcilability issue. We further optimized the schedule $s$ of the CTQW-inspired adiabatic path by an adaptive scheduling procedure. 
 
The modified CTQW-inspired adiabatic search simulation experiment demonstrates that indeed the $H_e$ without any modification leads to a constant time in the total running time, regardless of the search space size $N$. This result echoes the irreconcilability issue stated in \cite{wong2016irreconcilable}. On the other hand, the modified CTQW-inspired adiabatic path with catalyst Hamiltonian coefficient $\frac{\sin({s \pi})}{2}$ behaves similarly to the behavior of the optimal adiabatic local search. \black{Furthermore, the modifications are optimal and outperform the original adiabatic local search.} 

Lastly, in the adaptive adiabatic local search simulation with various coefficient functions experiment, we further investigated how to \black{reduce the time wasted in the sluggish period of an adiabatic local search path.} 
\black{As our numerical experiments show, the function $f_z^{\sine}(s)$ and $f_z^{grid}(s)$ provide significant improvement and both outperform the original adiabatic local search.} Even though 
the grid search $f_z^{grid}(s)$ approach could have further reduced the length of the sluggish (``slow down") interval, the benefit was offset by the additional  cost incurred from implementation over that of the other two methods. 

\begin{acknowledgments}
C.~C. gratefully acknowledges the support from the seed grant funding
(917035-13) from the State University of New York Polytechnic Institute and the
support from the Air Force Research Laboratory Summer Faculty Fellowship Program
(AFSFFP).
PMA would like to acknowledge support of this work from
the Air Force Office of Scientific Research (AFOSR). Any opinions, findings and conclusions or recommendations
expressed in this material are those of the author(s) and do not
necessarily reflect the views of Air Force Research Laboratory. 
The appearance of external hyperlinks does not constitute endorsement by the United States Department of Defense (DoD) of the linked websites, or the information, products, or services contained therein.  The DoD does not exercise any editorial, security, or other control over the information you may find at these locations.

\end{acknowledgments}

\appendix

\section{Time Integration of Adiabatic Local Search} \label{sect:sTot}
Given a spectral gap polynomial of the second order, that is  
\begin{equation}
g^2(s) = A (s^2 + bs + c) 
\end{equation} where $s$ is the Adiabatic schedule, and \footnote{this is the same as $g^2(t)$ as for each $t$ there is only corresponding $s$}
$\frac{ds}{dt} = \epsilon g^2(s)$, by integration on $t$ one obtains
\begin{eqnarray}
 T  = \int dt &=& \int_{0}^{1} \frac{ds} {\epsilon g^2(s)} =   \frac{1}{\epsilon A} \int_{0}^{1} \frac{ds}{(s^2 + bs + c)}.
\end{eqnarray}
\noindent 
(I) Case $b^2 - 4c > 0$: Let $r_{\pm} = \frac{-b\pm \sqrt{b^2 -4c}}{2}$. 
\begin{equation}
\int_{0}^{1} \frac{ds}{(s^2 + bs + c)} =    \frac{1}{r_+ - r_-} \int_{0}^{1} (\frac{1}{s-r_+}- \frac{1}{s-r_-})ds
\end{equation}
since $\int \frac{1}{s-a} ds = \ln |s-a|$.  Thus we have 
\begin{align}
T &= \frac{1}{\epsilon A (r_+ - r_-)} \ln \Big |\frac{s-r_+}{s-r_-}\Big|_0^1, \\
t &=  \frac{1}{\epsilon A (r_+ - r_-)} (\ln \Big |\frac{s-r_+}{s-r_-}\Big| - \ln \Big |\frac{r_+}{r_-}\Big | ).
\end{align} \\
(II) Case $b^2 -4c = 0$: 
\begin{equation}
\int_{0}^{1} \frac{ds}{(s^2 + bs + c)} =    \int_{0}^{1} \frac{1}{(s+b/2)^2}ds
\end{equation}
since $\int (s-a)^{-2} ds =-(s-a)^{-1}$, hence
\begin{align}
T&= \frac{-1}{\epsilon A}{\frac{1}{(s+(b/2))}\Big|_0^1} \\
t&= \frac{1}{\epsilon A}\Big ({\frac{s}{(b/2)(s+(b/2))}}\Big)
\end{align}
(III) Case $b^2 -4c <0$:
\begin{align}
\int_{0}^{1} \frac{ds}{(s^2 + bs + c)} 
&=\int_{0}^{1} \frac{1}{(s+b/2)^2 + \frac{4c-b^2}{4}}ds \\
&=\int_{b/2}^{1+(b/2)} \frac{1}{x^2 + (\sqrt{\frac{4c-b^2}{4}})^2}dx
\end{align}
since $\int\frac{1}{a^2 +x^2} dx =\frac{1}{a} \arctan \frac{x}{a}$. With $a =\sqrt{\frac{4c-b^2}{4}}$, we obtain 
\begin{align}
T&=\frac{1}{\epsilon A}(\frac{1}{a})(\arctan \frac{x}{a})\Big|_{b/2}^{1+(b/2)} \\
t&=\frac{1}{\epsilon A}(\frac{1}{a})(\arctan \frac{s+(b/2)}{a} -\arctan \frac{(b/2)}{a})
\end{align}
\newline 

\section{Energy Gap} \label{sect:simpleAmp}
Given an arbitrary 2 by 2 non-negative-entry Hermitian matrix $H$ as 
\begin{equation}
H = 
\begin{bmatrix}
\alpha & \gamma \\
\gamma & \beta \\
\end{bmatrix}, 
\end{equation}
via computing the determinant and eigenvalues, the energy gap $\Delta E$ is
\begin{equation}
\Delta E = | \lambda_+ - \lambda_-| = \sqrt{(\alpha - \beta )^2 + 4\gamma^2}.
\end{equation}
Simply from the view of energy gap, as long as $|\gamma|$ increases and the gap, $|\alpha -\beta|$, between the diagonal entries increases, the energy gap would increase. The increase of $|\gamma|$ can be 
adapted by $\sigma_x$ while $|\alpha -\beta|$ can be increased by $\sigma_z$.  They should be good candidates for the catalyst perturbation in the AQC path. Similarly, if the Hamiltonian has 
an imaginary part in the off diagonal entries, 
\begin{align}
H & = 
\begin{bmatrix}
\alpha & \gamma - di \\
\gamma+di & \beta \\
\end{bmatrix} \\
\Delta E &= | \lambda_+ - \lambda_-| = \sqrt{(\alpha -\beta)^2 + 4(\gamma^2+d^2)}.
\end{align}
The Hamiltonian $H$ (with no imaginery entries) can be expressed in terms of Pauli matrices as 
\begin{align}
H &= \frac{\alpha + \beta}{2} \mathbb{I} + \frac{\Delta E}{2}((\frac{2\gamma}{\Delta E})\sigma_x + ((\frac{\alpha -\beta}{2})(\frac{2}{\Delta E}) \sigma_z)) \\
 &=  \frac{\alpha + \beta}{2} \mathbb{I} + \frac{\Delta E}{2} A
\end{align}
such that, by use of power of Pauli matrices, 
\begin{equation}
e^{-iHt} = \cos(\frac{\Delta E t}{2})\mathbb{I} -i \sin(\frac{\Delta E t}{2})A.
\end{equation}

%

\end{document}